\RequirePackage{fix-cm}

\documentclass[pdftex,twocolumn,epjc3]{svjour3}          

\RequirePackage{graphicx}
\RequirePackage{mathptmx}      
\RequirePackage[colorlinks,citecolor=blue,urlcolor=blue,linkcolor=blue]{hyperref}
\RequirePackage{amsmath}
\RequirePackage{amssymb}
\RequirePackage{microtype}
\usepackage[absolute]{textpos}

\journalname{Eur. Phys. J. C}

\begin{document}

\title{Neutron-skin effect in direct-photon and charged-hadron production in Pb+Pb collisions at the LHC}

\author{Ilkka Helenius\thanksref{e1, addr1, addr2} \and
        Hannu Paukkunen\thanksref{e2, addr3, addr4, addr5} \and
        Kari~J. Eskola\thanksref{e3, addr3, addr4}}

\thankstext{e1}{email: ilkka.helenius@uni-tuebingen.de}
\thankstext{e2}{email: hannu.t.paukkunen@jyu.fi}
\thankstext{e3}{email: kari.eskola@jyu.fi}

\institute{Department of Astronomy and Theoretical Physics, Lund University, S\"{o}lvegatan 14A, SE-223 62 Lund, Sweden\label{addr1} \and 
\emph{Present Address:} Institute for Theoretical Physics, T\"{u}bingen University, Auf der Morgenstelle 14, 72076 T\"{u}bingen, Germany\label{addr2} \and 
University of Jyvaskyla, Department of Physics, P.O. Box 35, FI-40014 University of Jyvaskyla, Finland\label{addr3} \and 
Helsinki Institute of Physics, P.O. Box 64, FI-00014 University of Helsinki, Finland\label{addr4} \and 
Departamento de F\'{\i}sica de Part\'{\i}culas and IGFAE, Universidade de Santiago de Compostela, E-15782 Galicia, Spain\label{addr5}
}

\date{Received: date / Accepted: date}

\maketitle

\begin{textblock}{2}(12.35,0.3)
\mbox{}\hfill LU TP 16-33\\
\mbox{}\hfill June 2016
\end{textblock}

\begin{abstract}
A well-established observation in nuclear physics is that in neutron-rich spherical nuclei the distribution of neutrons extends farther than the distribution of protons. In this work, we scrutinize the influence of this so called neut\-ron-skin effect on the centrality dependence of 
high-$p_{\rm T}$ direct-photon and charged-hadron production. We find that due to the estimated spatial dependence of the nuclear parton distribution functions, it will be demanding to unambiguously expose the neutron-skin effect with direct photons. However, when taking a ratio between the cross sections for negatively and positively charged high-$p_{\rm T}$ hadrons, even centrality-dependent nuclear-PDF effects cancel, making this observable a better handle on the neutron skin. Up to 10~\% effects can be expected for the most peripheral collisions in the measurable region.
\end{abstract}

\section{Introduction}

In ultra-relativistic heavy-ion collisions the concept of centrality plays an important role in phenomena such as the jet energy loss \cite{Abelev:2013kqa,Chatrchyan:2012nia,Aad:2014bxa} or the systematics of azimuthal anisotro\-pies \cite{Aad:2014eoa,Chatrchyan:2012xq}. Experimentally, the centrality of a collision is usually defined according to the amount of energy seen in a specific part of the detector, typically at large pseudorapidities \cite{Aad:2014eoa,Chatrchyan:2011sx,Abelev:2013qoq}: the more energy observed, the more central the collision. The theoretical centrality categorizations are based on Glauber models \cite{Miller:2007ri}, in which the centrality is related to impact parameter (optical Glauber) or to the number of nucleon--nucleon collisions (Monte-Carlo Glauber). While there is no direct, unambiguous relation between the experimental and theoretical prescriptions, it is yet generally accepted that a correspondence exists in collisions of two heavy nuclei. In nucleon--nucleus collisions, however, the same experimental procedure has led to rather unexpected results \cite{Chatrchyan:2014hqa,ATLAS:2014cpa,Adare:2015gla} and it is now commonly believed that such a centrality classification induces a non-trivial bias on the hard process whose centrality dependence was to be measured \cite{Martinez-Garcia:2014ada,Bzdak:2014rca,Alvioli:2014eda,Perepelitsa:2014yta,Armesto:2015kwa}.

The Glauber models take the nuclear density distribution as an input and it is typically assumed to be identical for protons and neutrons. However, the measurements at lower energies indicate that the tail of the neutron density distribution extends farther than that of the proton density \cite{Tarbert:2013jze,Tsang:2012se,Zenihiro:2010zz}. While this so-called neutron-skin (NS) effect \cite{Horowitz:2000xj} should not have a great importance in the centrality classification itself, it leads to a growth of the relative number of neutrons at high impact parameters and thereby influences the observables sensitive to electroweak effects in peripheral (large impact parameter) collisions of two heavy nuclei. The impact of the NS effect to $W^{\pm}$ production in Pb+Pb and p+Pb collisions at the LHC was studied in Ref.~\cite{Paukkunen:2015bwa}. 

In this work, we extend the study of Ref.~\cite{Paukkunen:2015bwa} to direct-photon and charged-hadron production at high transverse momenta ($p_{\rm T}$) in Pb+Pb collisions at the LHC. The goal is to study whether the NS effect has a measurable impact on these observables and to quantify at which centralities and kinematics (transverse momentum, rapidity) the effect would be most pronounced. Our hope is that, later on, the NS effect could help to calibrate the centrality classification in collisions involving heavy ions.

\section{Centrality-dependent hard-process cross section}

Centrality classification is done here using the optical Glau\-ber model as in Refs.~\cite{Paukkunen:2015bwa,Helenius:2012wd}. For the nuclear density distribution we use the two-parameter Fermi (2pF) distribution,
\begin{equation}
\rho^{A}(\mathbf{r}) = \rho_0^{A}/(1 + \mathrm{e}^{(|\mathbf{r}| - d_{A})/a_{A}}),
\end{equation}
where $d_{A}$ describes the radius of the nucleus and $a_{A}$ the thickness of the nuclear surface (skin) in nucleus with a mass number $A$. To account for the NS effect the nuclear density is written as 
$
\rho^{A}(\mathbf{r}) = \rho^{\mathrm{p},A}(\mathbf{r}) + \rho^{\mathrm{n},A}(\mathbf{r}) 
$
where now the parameters of the 2pF distribution are different for protons and neutrons. Here we use the parameters from Ref.~\cite{Tarbert:2013jze}, $d_\mathrm{n, Pb} = 6.70 \pm 0.03~\mathrm{fm}$ and $a_\mathrm{n, Pb} = 0.55 \pm 0.03~\mathrm{fm}$, for neutrons and $d_\mathrm{p, Pb} = 6.680~\mathrm{fm}$ and $a_\mathrm{p, Pb} = 0.447~\mathrm{fm}$ for protons.\footnote{In the analysis of Ref.~\cite{Tarbert:2013jze}, the proton density was taken as fixed when fitting the neutron parameters to the data. Therefore, the proton density has no uncertainty here.}

The hard-process cross section in an $A+B$ collision for a given centrality class $\mathcal{C}_k$ corresponding to an impact parameter interval $b_k\leq b < b_{k+1}$ (where $b=|\mathbf{b}|$) can be calculated from
\begin{align}
\mathrm{d}\sigma_{AB}^{\mathrm{hard}}(\mathcal{C}_k) =& \,2\pi \int_{b_{k}}^{b_{k+1}} \mathrm{d}b\, b \int \mathrm{d}^2\mathbf{s} \sum_{i,j} T_A^i(\mathbf{s_1})\,T_B^j(\mathbf{s_2}) \notag \\ &\mathrm{d}\sigma_{ij}^{\mathrm{hard}}(A,B,\mathbf{s_1}, \mathbf{s_2}),
\label{eq:masterformula}
\end{align}
where the nuclear thickness functions $T_A^i(\mathbf{s})$ are obtained by integrating the density over the longitudinal (i.e. beam) direction,
\begin{equation}
T^i_{A}({\bf s}) \equiv \int_{-\infty}^{\infty} \mathrm{d}z \rho^{{\rm i},A}({\bf r}),
\end{equation}
and $\mathbf{s_{1,2}} = \mathbf{s} \pm \mathbf{b}/2$ are defined according to Fig. 20 of Ref.~\cite{Helenius:2012wd}. The indices $i$ and $j$ run over combinations $(i,j) = (\mathrm{p},\mathrm{p}),\, (\mathrm{p},\mathrm{n}),$\\$\, (\mathrm{n},\mathrm{p})\,\text{and}\,(\mathrm{n},\mathrm{n}) $. The impact-parameter in\-ter\-vals required in Eq.~(\ref{eq:masterformula}) correspond to the fractions of the total inelastic cross section $\sigma^{\rm inel}_{AB}(\sqrt{s})$, obtained as in Refs.~\cite{Paukkunen:2015bwa,Helenius:2012wd} by
\begin{equation}
\sigma^{\rm inel}_{AB}(\sqrt{s}) = \int_{-\infty}^{\infty} \mathrm{d}^2{\bf b} \left[ 1 - \mathrm{e}^{-T_{AB}({\bf b}) \, \sigma^{\rm inel}(\sqrt{s})}\right],
\end{equation}
where
\begin{equation}
T_{AB}({\bf b}) \equiv \int_{-\infty}^{\infty} \mathrm{d}^2{\bf s}
\sum_{i,j}T_A^{j}(\mathbf{s_1})\, T_B^{i}(\mathbf{s_2}).
\end{equation}
We take $\sigma^{\rm inel}({\sqrt{s}=5~\mathrm{TeV}})=70~\mathrm {mb}$ \cite{Antchev:2013iaa}. The spatial dependence of the hard-process cross section $\mathrm{d}\sigma_{ij}^{\mathrm{hard}}$ arises here from the spatial dependence of nPDFs,
\begin{multline}
\mathrm{d}\sigma_{ij}^{\mathrm{hard}}(A,B,\mathbf{s_1}, \mathbf{s_2}) = \sum_{k,l}f_k^{i/A}(x_1, Q^2, \mathbf{s_1})\\ \otimes f_l^{j/B}(x_2, Q^2, \mathbf{s_2}) \otimes\mathrm{d}\hat{\sigma}^{kl\rightarrow {\rm observable}}, \label{eq:hxsec}
\end{multline}
where $\mathrm{d}\hat{\sigma}^{kl\rightarrow {\rm observable}}$ are perturbative coefficient functions and $k$ and $l$ are parton flavour indices. The nPDFs appearing in Eq.~(\ref{eq:hxsec}) above are defined as
\begin{equation}
f_k^{i/A}(x, Q^2, \mathbf{s}) = r_k^{i/A}(x,Q^2,\mathbf{s})f_k^{i}(x, Q^2), 
\end{equation}
where $f_k^{i}(x, Q^2)$ is the free nucleon PDF (here CT10NLO \cite{Lai:2010vv}) and $r_k^{i/A}(x,Q^2,\mathbf{s})$ the nuclear modification which depends on the transverse position of the nucleon inside the nucleus.\footnote{Currently, there is no coherent way to treat the PDF nuclear modifications within the Monte-Carlo Glauber model. This is actually why, in this work, we stick to the optical version of the Glauber model.} Here we use \textsc{EPS09s} nuclear modifications from Ref.~\cite{Helenius:2012wd} in which
\begin{equation}
 r_k^{{\rm p}/A}(x,Q^2,\mathbf{s}) = 1 + \sum_{j=1}^4 c^{j}_{k}(x,Q^2) \left[T_A^{\rm p}(\mathbf{s}) + T_A^{\rm n}(\mathbf{s})\right]^j,
 \label{eq:eps09s}
\end{equation}
where the coefficients $c^{j}_{k}(x,Q^2)$ are obtained by analyzing the $A$-dependence of the \textsc{EPS09} \cite{Eskola:2009uj} nPDFs. The neutron PDFs $f_k^{{\rm n}/A}(x, Q^2, \mathbf{s})$ are obtained from the proton PDFs $f_k^{{\rm p}/A}(x, Q^2, \mathbf{s})$ by the isospin symmetry. By combining all, Eq.~(\ref{eq:masterformula}) factorizes into purely geometric and purely momentum-depen\-dent parts which can be evaluated separately thereby reducing the dimensions of the required numerical integrations. We use the \textsc{Incnlo} program \cite{Aurenche:1987fs,Aversa:1988vb,Aurenche:1998gv,Aurenche:1999nz} to calculate the momentum-dependent parts at next-to-leading order in perturbative QCD. 

\section{Results}

\subsection{Direct-photon production}

\begin{figure*}[htb!]
\centering
\includegraphics[width=0.49\textwidth]{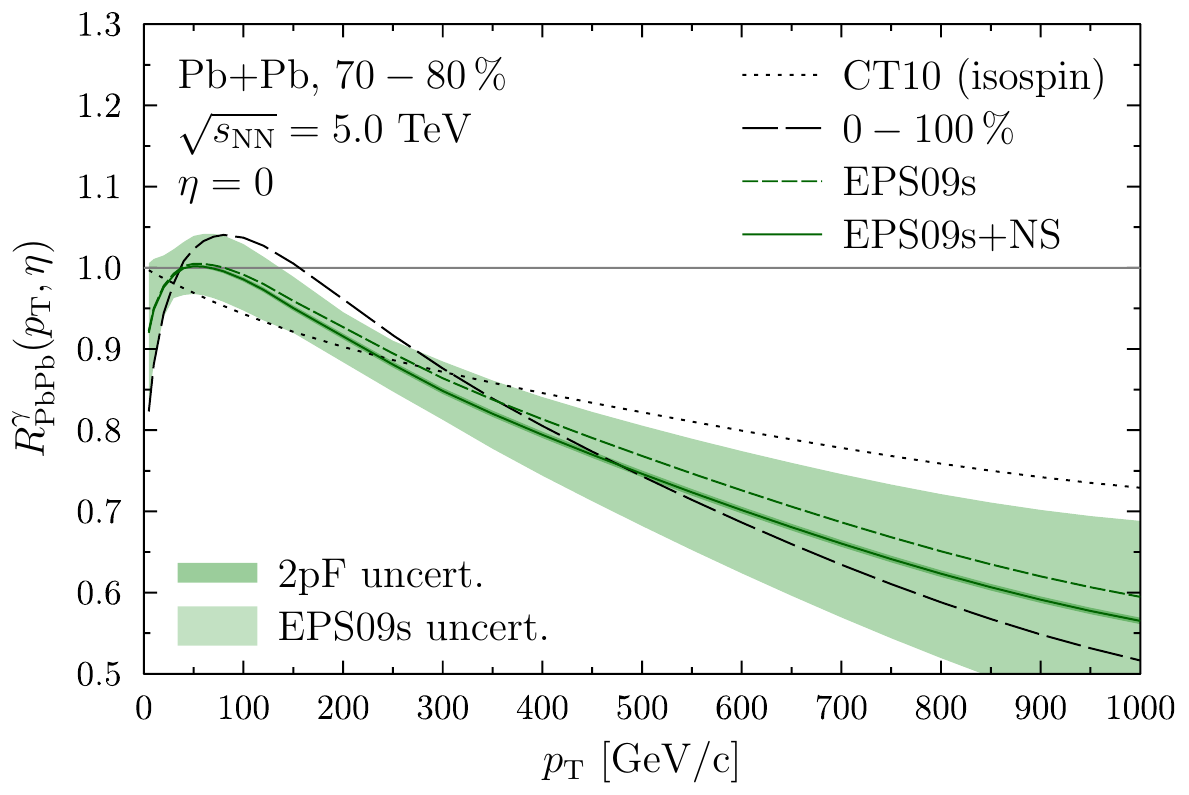}
\includegraphics[width=0.49\textwidth]{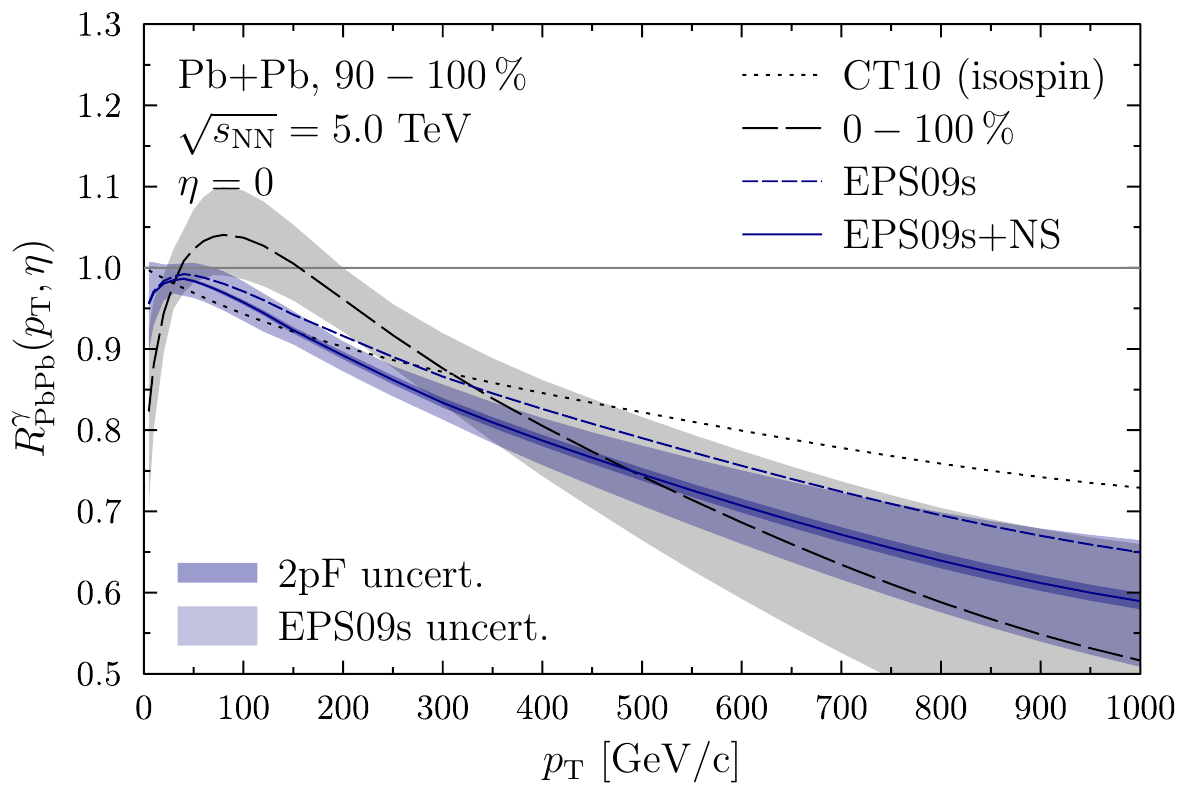}
\caption{The nuclear modification factor for direct-photon production in peripheral Pb+Pb collisions with $\sqrt{s_{\mathrm{NN}}} = 5.0~\mathrm{TeV}$ at mid-rapidity for two centralities 70--80~\% (left) and 90--100~\% (right). The results are compared to the isospin effect (dotted) and MB result (long-dashed) and the centrality-dependent results are shown with (solid) and without (short-dashed) the NS effect. The uncertainties from \textsc{EPS09s} nPDFs (light colour band) and 2pF parametrization (dark colour band around the solid lines) are calculated with the NS effect. The gray band shows the \textsc{EPS09s} uncertainty for the MB result in the right-hand panel.}
\label{fig:RPbPbgamma}
\end{figure*}

Direct photons are produced either in the hard process or by the fragmentation of high-$p_{\rm T}$ partons from the hard process. To obtain the latter contribution we convolute the partonic spectra with the BFG (set II) parton-to-photon fragmentation functions (FFs) \cite{Bourhis:1997yu}. Since the photon coupling is stronger in the case of up-type quarks than with the down-type quarks, the production rate of direct photons is larger in p+p collisions than in n+n collisions. This leads to a lower per-nucleon rate of direct photons in heavy-ion collisions than in p+p collisions due to the presence of neutrons. This is often referred to as the isospin effect and it becomes important at large values of $x$ where the valence quarks dominate. Furthermore, since the relative fraction of neutrons grows towards the edge of nucleus due to the NS effect, an additional suppression of direct photons in peripheral collisions is expected. 

\begin{figure*}[thb!]
\centering
\includegraphics[width=0.49\textwidth]{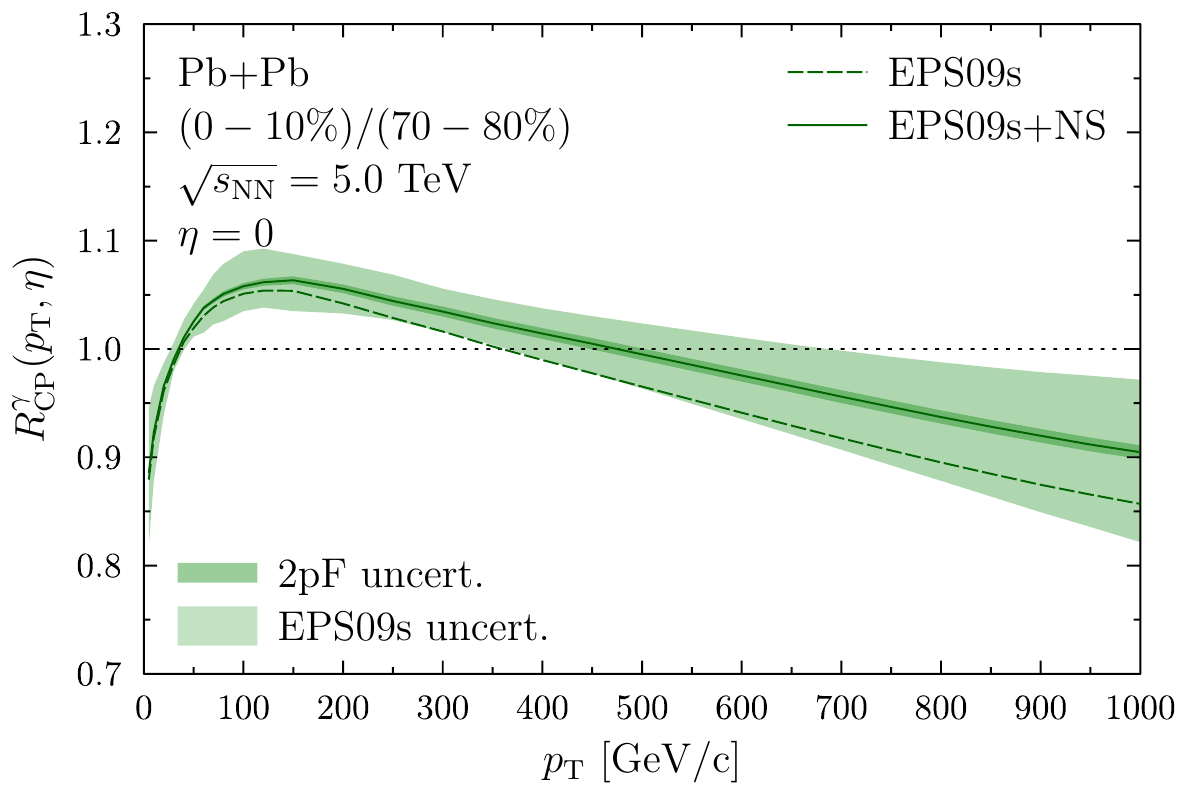}
\includegraphics[width=0.49\textwidth]{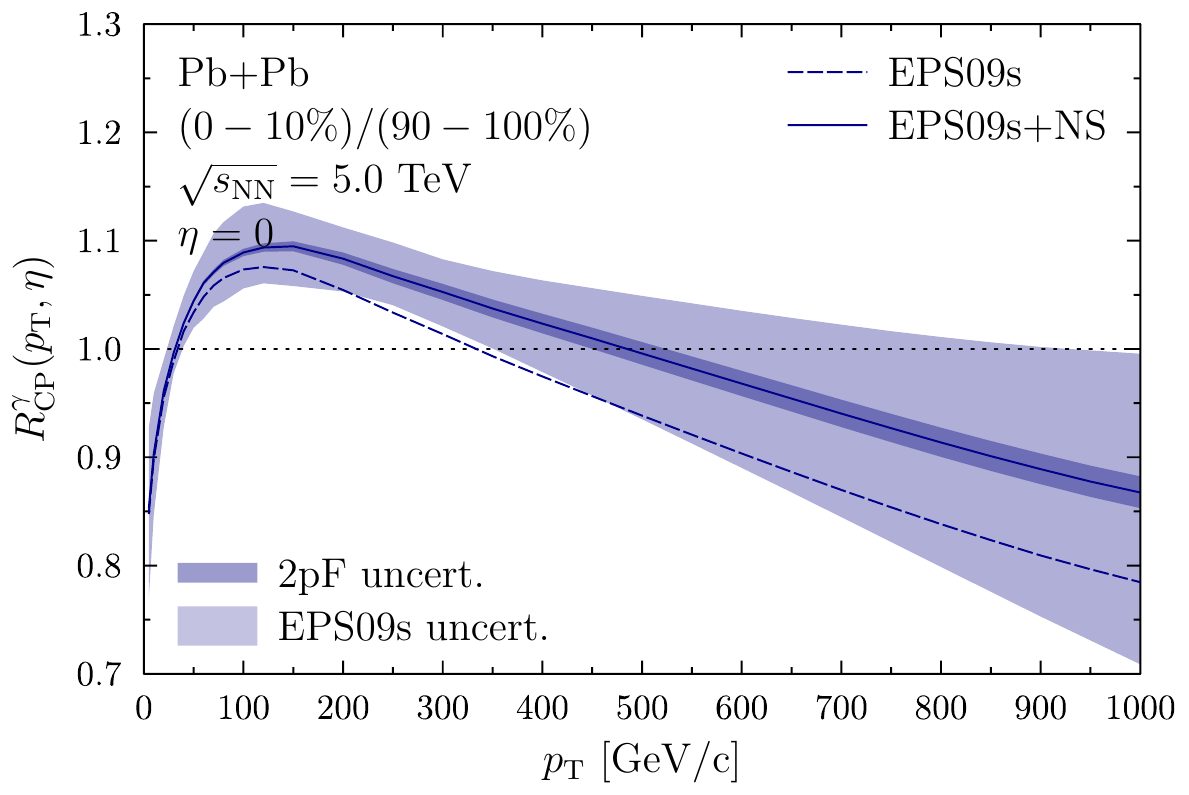}
\includegraphics[width=0.49\textwidth]{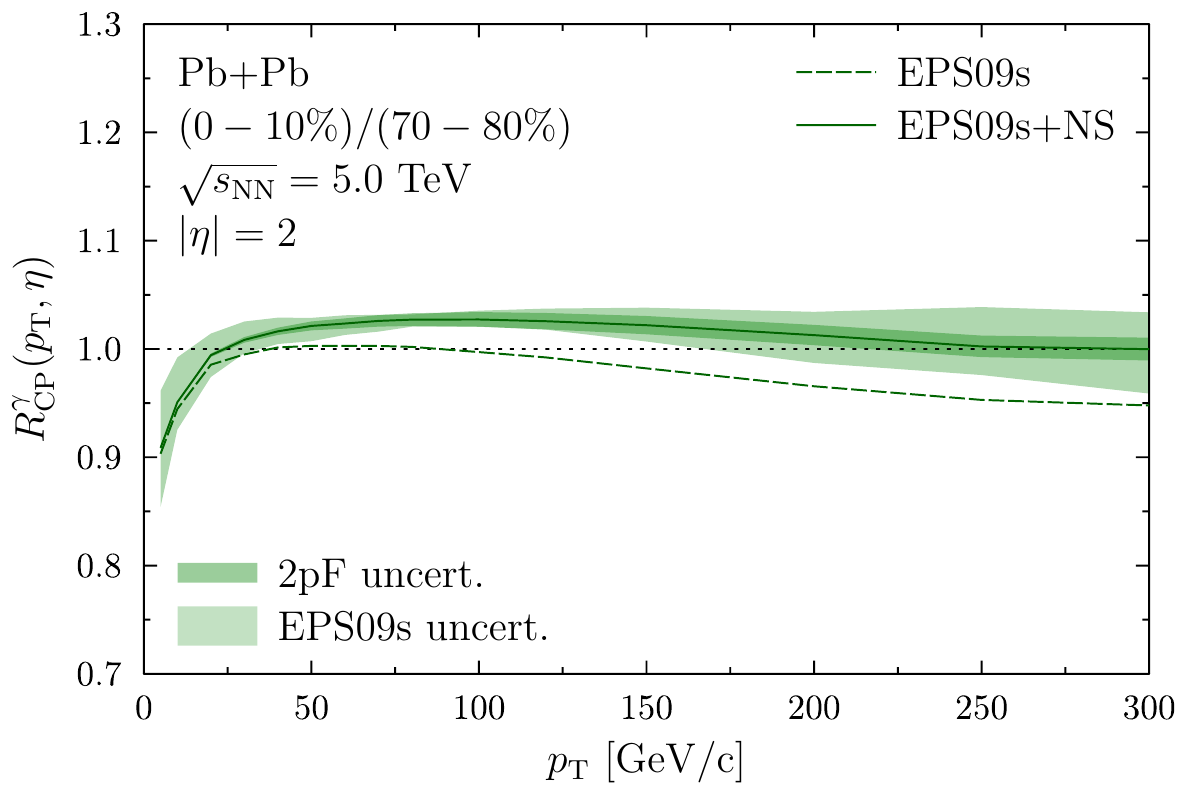}
\includegraphics[width=0.49\textwidth]{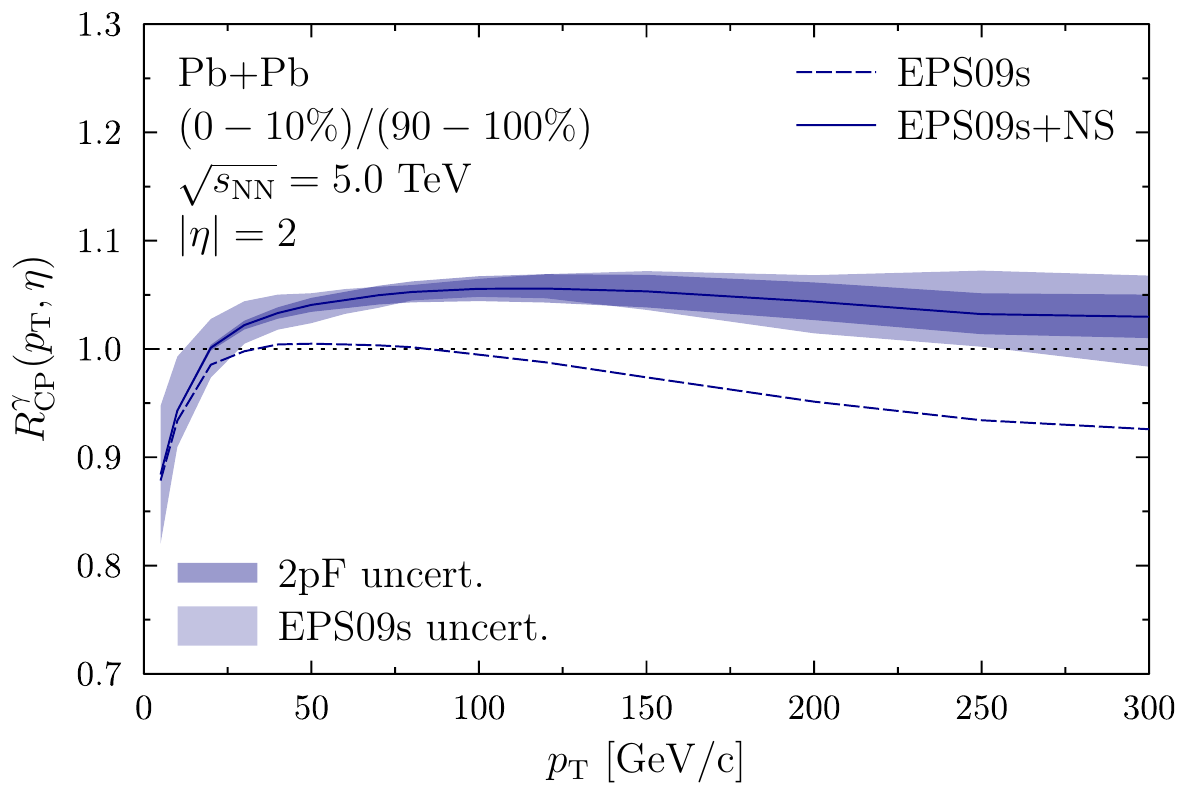}
\caption{The central-to-peripheral ratio for direct-photon production in Pb+Pb collisions with $\sqrt{s}=5.0~\mathrm{TeV}$ at mid-rapidity for (0--10~\%)/(70--80~\%) (upper left) and (0--10~\%)/(90--100~\%) (upper right), and at forward rapidity for (0--10~\%)/(70--80~\%) (lower left) and (0--10~\%)/(90--100~\%) (lower right). The results with (solid) and without (dashed) NS effects are shown and the uncertainties are from the \textsc{EPS09s} nPDFs (light colour band) and from the 2pF-parametrization (dark colour bands).}
\label{fig:RCPgamma}
\end{figure*}

A canonical way to quantify the nuclear effects is to compute the nuclear modification factor, defined in a given centrality class $\mathcal{C}_k$ of a Pb+Pb collision as
\begin{equation}
R_{\mathrm{PbPb}}^{\gamma}(\mathcal{C}_k) = \frac{1}{T_{\mathrm{PbPb}}(\mathcal{C}_k)}\frac{\mathrm{d}\sigma_{\mathrm{PbPb}}^{\gamma}(\mathcal{C}_k)}{\mathrm{d}p_{\mathrm{T}}\mathrm{d}\eta} \bigg/ \frac{\mathrm{d}\sigma_{\mathrm{pp}}^{\gamma}}{\mathrm{d}p_{\mathrm{T}}\mathrm{d}\eta}, 
\end{equation}
where the normalization is related to the amount of interacting nuclear matter, 
\begin{equation}
T_{\mathrm{PbPb}}(\mathcal{C}_k) = 2 \pi \int_{b_k}^{b_{k+1}}\mathrm{d}b\,b\int\mathrm{d}^2\mathbf{s}\, \sum_{i,j}T_{\rm Pb}^{j}(\mathbf{s_1})\, T_{\rm Pb}^{i}(\mathbf{s_2}),
\end{equation}
where the impact parameters $b_k$ and $b_{k+1}$ define the centrality class $\mathcal{C}_k$ as in Eq.~(\ref{eq:masterformula}). For the cross-section calculations we have set the renormalization, factorization and fragmentation scales to photon $p_{\rm T}$. The uncertainties related to the scale ambiguities are not considered here in more detail since they largely cancel out in the ratio, especially at large values of $p_{\rm T}$ relevant here \cite{Helenius:2013bya}. The isolation criterion, often used by experiments to suppress secondary photons from hadronic decays, is not applied here since the effect to $R_{\mathrm{PbPb}}^{\gamma}$ is negligible at the very high values of $p_{\mathrm{T}}$ considered here.\footnote{The valence quark-gluon channel dominates irrespectively of the isolation.}

Figure~\ref{fig:RPbPbgamma} shows $R_{\mathrm{PbPb}}^{\gamma}$ at mid-rapidity for two centrality classes, 70--80~\% and 90--100~\% with and without the NS effect, compared also to the minimum-bias (0--100~\%, MB) result and to the isospin effect. The uncertainties considered here are the \textsc{EPS09s} uncertainty (light colour band) and the one related to the uncertainty of neutrons 2pF parameters (dark colour band), obtained by evaluating $R_{\mathrm{PbPb}}^{\gamma}$ with the quoted parameter variations, and adding the differences to the central prediction in quadrature. 

The different $p_{\rm T}$ regions are sensitive to different nPDF effects. First, comparing the MB result to the result without nPDF effects (only isospin), at $p_{\rm T} < 30~\mathrm{GeV/c}$ some suppression due to shadowing is observed which then turns into an enhancement due to anti-shadowing. At $p_{\rm T} > 300~\mathrm{GeV/c}$ a suppression due to the EMC effect is observed. The spatial dependence of the nPDFs always decreases the nuclear effects towards more peripheral collisions whereas the NS effect generates additional suppression with increasing $p_{\rm T}$. Therefore, at high values of $p_{\rm T}$ where the impact of NS is more pronounced, these two effects pull towards opposite directions thereby ``softening'' the aggregate centrality dependence.

As the nuclear modifications of the PDFs gradually disappear with increasing peripherality, also the uncertainty becomes smaller for more peripheral events. However, even in the 90--100~\% bin the nPDF uncertainty is of the same order as the NS effect which further complicates the separation of different effects. The non-zero nPDF effects even at the most peripheral bin are due to the power series ansatz in \textsc{EPS09s}, see Eq.~(\ref{eq:eps09s}), which, by construction, gives zero nuclear modifications only when $b\rightarrow \infty$. The uncertainty from the neutron 2pF parametrization turns out to be rather small as the contribution of the n+n channel is inferior e.g. to the contribution of the p+p channel and thus the variations in the neutron density are not that important. Since the fraction of neutrons (and therefore the n+n channel contribution) grows towards more peripheral collisions, also the uncertainty grows accordingly.

The centrality dependence can also be studied using the central-to-peripheral ratio $R_{\rm CP}$ defined as
\begin{equation}
R_{\rm CP} = \frac{T_{\mathrm{PbPb}}(P)}{ T_{\mathrm{PbPb}}(C)}\frac{\mathrm{d}\sigma_{\mathrm{PbPb}}^{\gamma}(C)}{\mathrm{d}p_{\mathrm{T}}\mathrm{d}\eta} \bigg/ \frac{\mathrm{d}\sigma_{\mathrm{PbPb}}^{\gamma}(P)}{\mathrm{d}p_{\mathrm{T}}\mathrm{d}\eta}.
\end{equation}
The advantage is that there is no need for a separate p+p baseline measurement and also that some uncertainties are expected to cancel out. Here, we have used the bin 0--10~\% as the central result and compared it to the 70--80~\% and 90--100~\% bins. The results are shown in Fig.~\ref{fig:RCPgamma} again with and without the NS effect. Since the peripheral bins are now in the denominator, the NS effect increases the ratio and therefore decreases the centrality dependence at high-$p_{\rm T}$ region ($R_{\rm CP}$ closer to unity). The nPDF originating uncertainties are now larger with the most peripheral bin red (90--100~\%) because the uncertainties in the central bin (similar to the MB uncertainty in Fig.~\ref{fig:RPbPbgamma}) do not cancel here as effectively as with the less peripheral bin (70--80~\%). Even though the interpretation of this observable is easier, the NS effect is still of the same order as the nPDF uncertainties. 

At forward/backward rapidities (the lower panels in Fig.~\ref{fig:RCPgamma}) the nPDF uncertainties are smaller. This is because here the dominant contribution comes from $q$+g initial state where the gluon is at shadowing region with only mild uncertainty (at high factorization scale), and the quark is also a well-constrained high-$x$ valence quark. The modifications, however, are quite small and since there is an additional uncertainty due to modelling of the spatial dependence of the nPDFs, it is difficult to unambiguously study the NS effect with this observable. The most accurate centrality-dependent measurement for photons in Pb+Pb comes from the ATLAS collaboration \cite{Aad:2015lcb}. However, their most peripheral bin 40--80~\% is still too central, and also the experimental uncertainties are large, to see any effects of NS. 

\subsection{Charged-hadron production}

\begin{figure*}[thb!]
\centering
\includegraphics[width=0.49\textwidth]{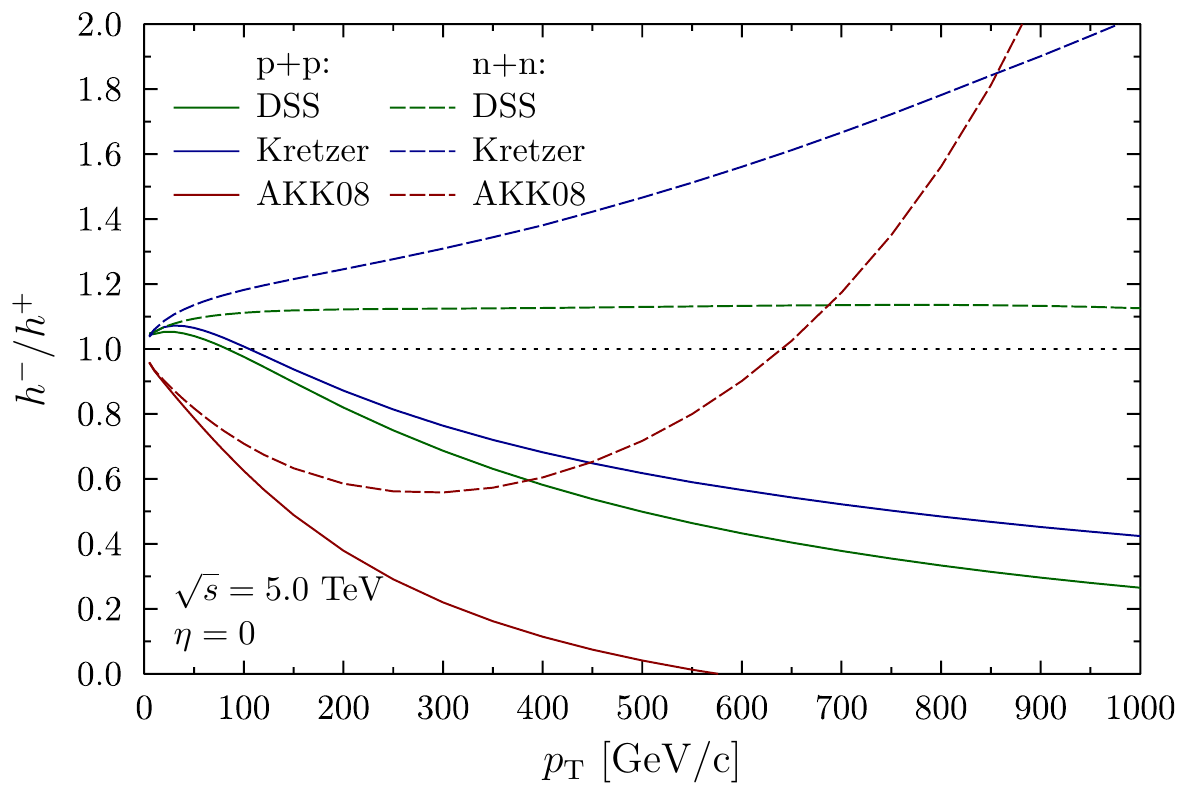}
\includegraphics[width=0.49\textwidth]{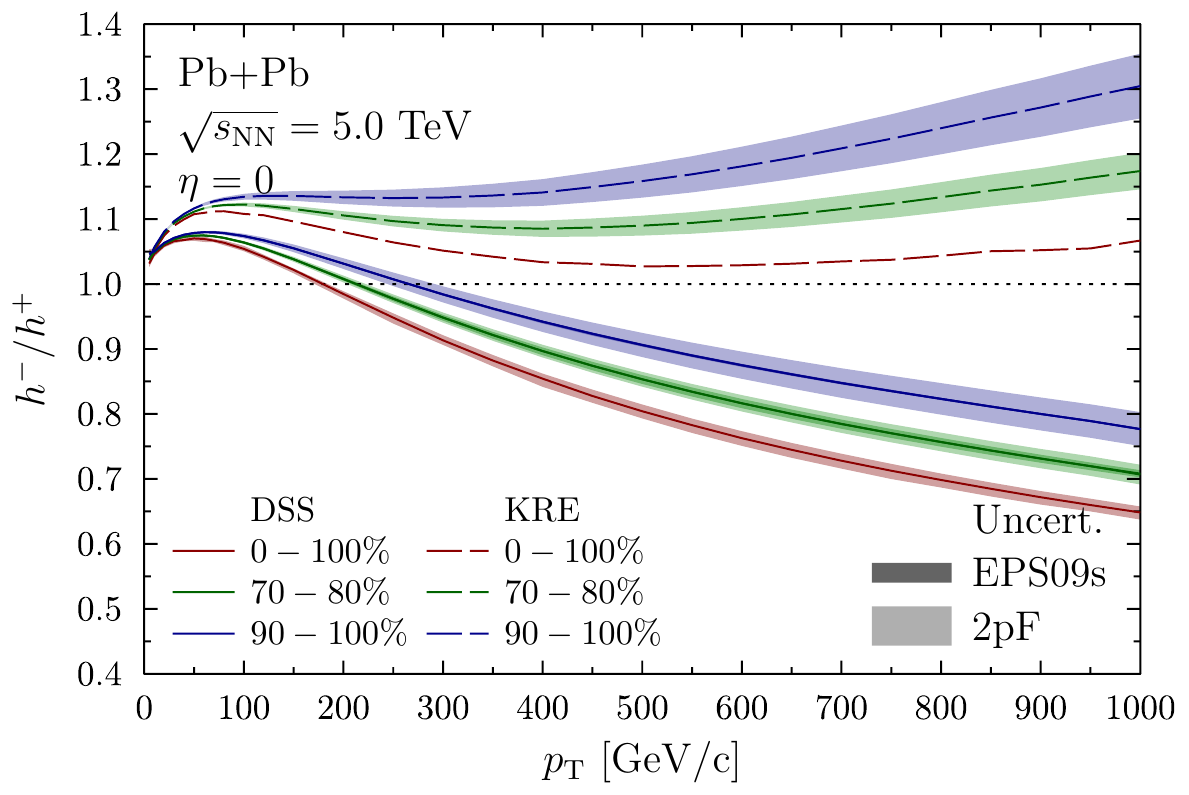}
\includegraphics[width=0.49\textwidth]{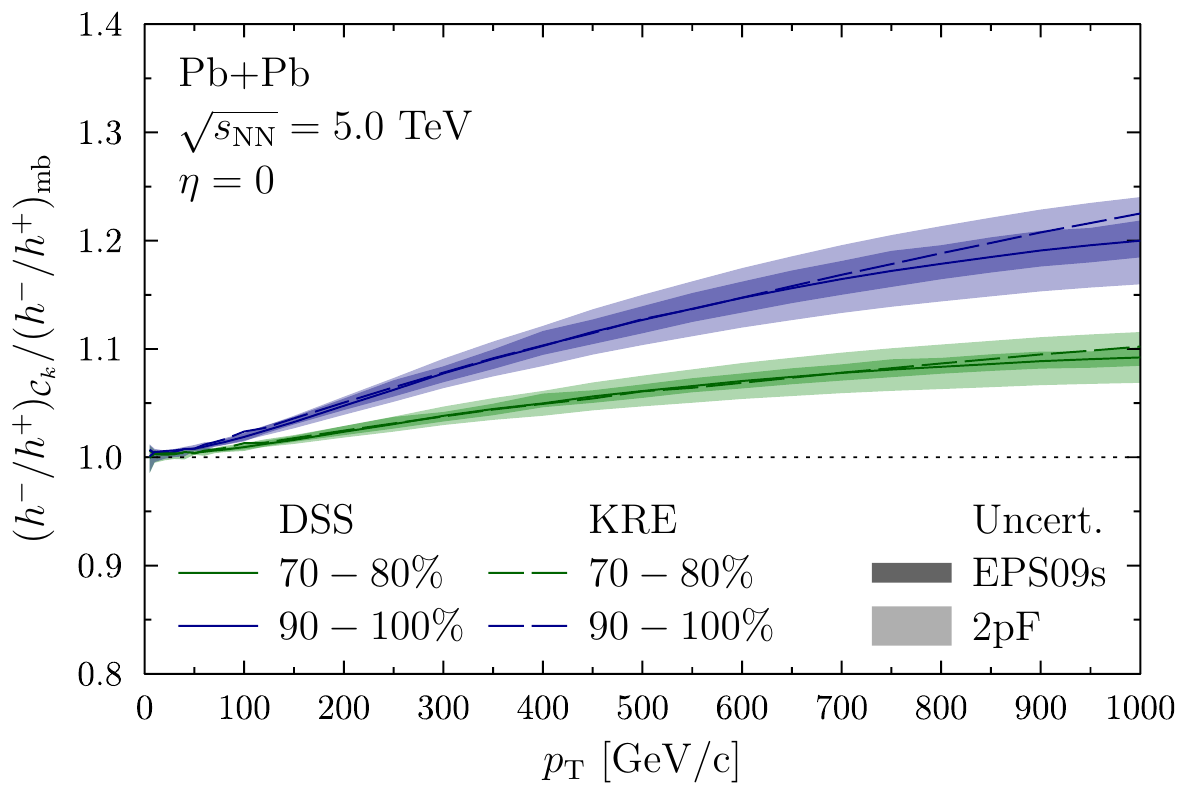}
\includegraphics[width=0.49\textwidth]{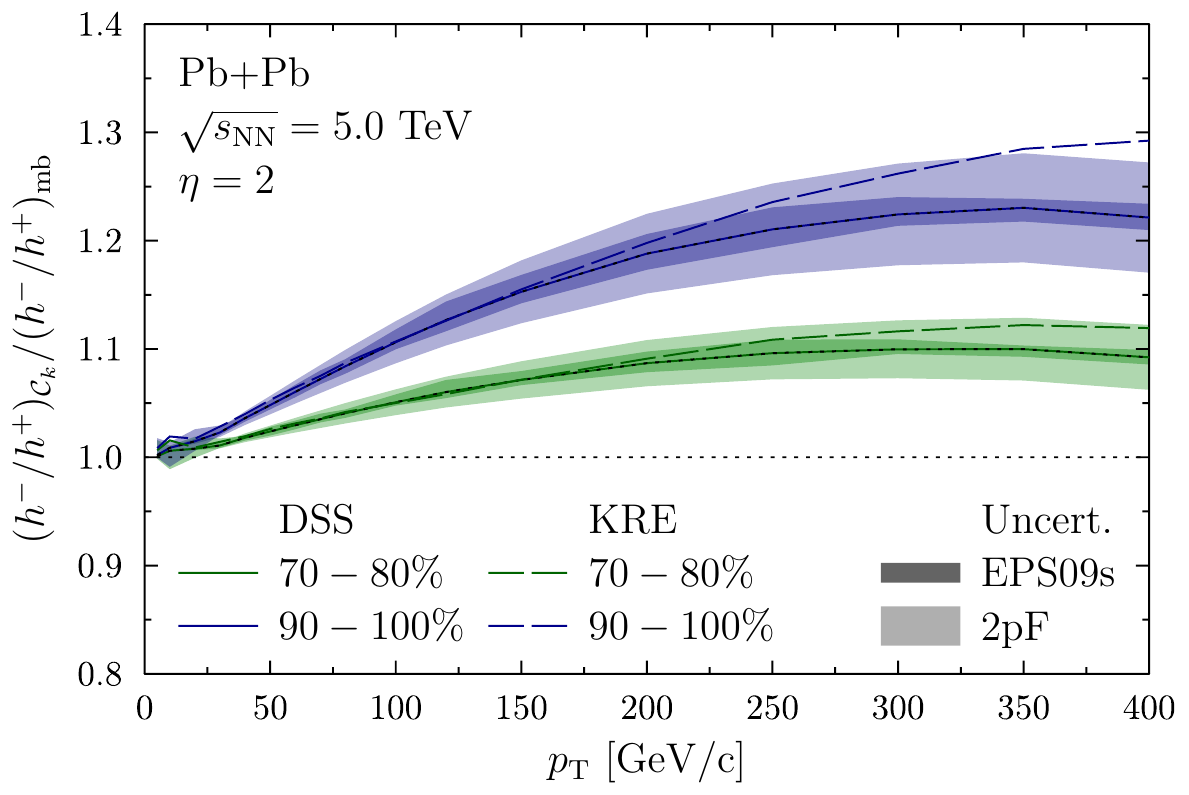}
\caption{\textbf{Upper left panel:} The ratio between positively and negatively charged hadrons in p+p (solid) and in n+n (dashed) collisions at $\sqrt{s}=5.0~\mathrm{TeV}$ and $\eta=0$ using DSS (green), Kretzer (blue), and AKK08 (red) FFs. \textbf{Upper right panel:} The ratio between positively and negatively charged hadrons in Pb+Pb collisions at $\sqrt{s}=5.0~\mathrm{TeV}$ and $\eta=0$ using DSS (solid) and Kretzer (dashed) FFs for centrality classes 0--100~\% (red), 70--80~\% (green), and 90--100~\% (blue). The light colour bands show the uncertainty from the 2pF parametrization and the dark one the EPS09s uncertainty with DSS. For the 0--100~\% (90--100~\%) bin only the EPS09s (2pF) uncertainty is visible. \textbf{Lower panels:} The ratio between positively and negatively charged hadrons in Pb+Pb collisions at $\sqrt{s}=5.0~\mathrm{TeV}$ and $\eta=0$ in 70--80~\% (green) and 90--100~\% (blue) centralities normalized with 0--100~\% using DSS (solid) and Kretzer (dashed) FFs. The colour bands show the nPDF uncertainties (dark) and uncertainty in the 2pF parametrization (light) with DSS FFs.}
\label{fig:RnegPosHadr}
\end{figure*}

An observable in which the nPDF effects should cancel out very efficiently but yet be sensitive to the NS effect, is the ratio between negatively and positively charged hadrons,
\begin{equation}
\frac{h^-}{h^+}(\mathcal{C}_k) = \frac{\mathrm{d}\sigma_{\mathrm{PbPb}}^{h^-}(\mathcal{C}_k)}{\mathrm{d}p_{\mathrm{T}}\mathrm{d}\eta} \bigg/\frac{\mathrm{d}\sigma_{\mathrm{PbPb}}^{h^+}(\mathcal{C}_k)}{\mathrm{d}p_{\mathrm{T}}\mathrm{d}\eta}.
\label{eq:Rhposneg}
\end{equation}
Since the relative number of neutron-involving (p+n, n+p, n+n) collisions is higher in peripheral than in central collisions, the increased d-quark contribution produces less positively charged hadrons and more negatively charged hadrons during the fragmentation. Here we do not consider any additional final-state effects that may affect the hadron production even though a significant suppression for the production of high-$p_{\mathrm{T}}$ hadrons has been observed \cite{Abelev:2014laa,Adam:2015kca,Khachatryan:2016odn} in all centralities. Indeed, the measurements in Refs.~\cite{Abelev:2014laa,Adam:2015kca} show that the suppression at high $p_{\rm T}$ ($p_{\mathrm{T}} \gtrsim 10~\mathrm{GeV/c}$) is very similar for all light charged hadrons (pions, kaons, protons) and, consequently, the particle ratios $(\mathrm{K}^++\mathrm{K}^-)/(\pi^++\pi^-)$ and $({\rm p}+\overline{{\rm p}})/(\pi^++\pi^-)$ are the same in p+p and Pb+Pb collisions. This motivates us to conjecture that final-state effects would have only a relatively small influence on the ratio of Eq.~(\ref{eq:Rhposneg}). Moreover, at the very high-$p_{\mathrm{T}}$ region ($p_{\rm T} \gg 100~\mathrm{GeV/c}$) where the current measurements are still statistically limited \cite{Khachatryan:2016odn}, the suppression effect in peripheral bins may be even negligible.

The cross section for hadron production is calculated by convoluting the partonic spectra with non-perturbative parton-to-hadron FFs. We consider three options, \textsc{dss} \cite{deFlorian:2007ekg}, \textsc{kretzer} \cite{Kretzer:2000yf} and \textsc{akk08} \cite{Albino:2008fy}. To better understand the variations seen using different FFs, the $h^-/h^+$ ratios in p+p and n+n collisions at $\sqrt{s} = 5.0~\mathrm{TeV}$ are shown in Fig.~\ref{fig:RnegPosHadr}. The first observation is that with \textsc{akk08} FFs the ratio in p+p actually turns negative at high-$p_{\mathrm{T}}$, caused by the cross section for $h^-$ becoming negative. This clearly unphysical result implies that the considered kinematic region is out of the validity region of \textsc{akk08}. The results using \textsc{dss} and \textsc{kretzer} are not that different in p+p collisions but for n+n collisions almost a factor of two difference at the very highest values of $p_{\rm T}$ is observed. These differences between the FF analyses generate some further theoretical uncertainty for the considered observable. Turning this around, a measurement of $h^-/h^+$ in p+Pb or Pb+Pb collisions would clearly provide additional constraints for future FF analyses (modulo the possible final-state effects in Pb+Pb).

The $h^-/h^+$ ratios in Pb+Pb collisions at $\sqrt{s_{\mathrm{NN}}} = 5.0~\mathrm{TeV}$ for centrality classes 70--80~\% and 90--100~\% are shown in the upper right panel of Fig.~\ref{fig:RnegPosHadr} together with the MB result with \textsc{dss} and \textsc{kretzer} FFs. The nPDF effects, including the centrality dependence and the uncertainties, are found to be negligible as expected. The uncertainty in the 2pF parametrization is negligible for the MB case but increases towards more peripheral collisions. The uncertainties are larger than in the case of direct-photon production as the cross section for $h^-$ now gets a large contribution from the n+n channel and thus carries more sensitivity to the parameter uncertainties in the neutron density. More importantly, the centrality dependence from the NS effect is clearly visible in this observable. However, the different FFs still yield rather different results but normalizing the ratio with the MB result, the FF dependence largely cancels out. This is demonstrated in the lower panels of Fig.~\ref{fig:RnegPosHadr}, where the ratios in 70--80~\% and 90--100~\% classes are normalized with the 0--100~\% result for $\eta = 0$ and $|\eta| = 2$. Some FF dependence persists with $|\eta| = 2$ but it is still smaller or of the same order than the uncertainty in 2pF parametrization. Also the nPDFs yield a few-percent uncertainty for the observable.

\begin{figure}[thb!]
\centering
\includegraphics[width=\linewidth]{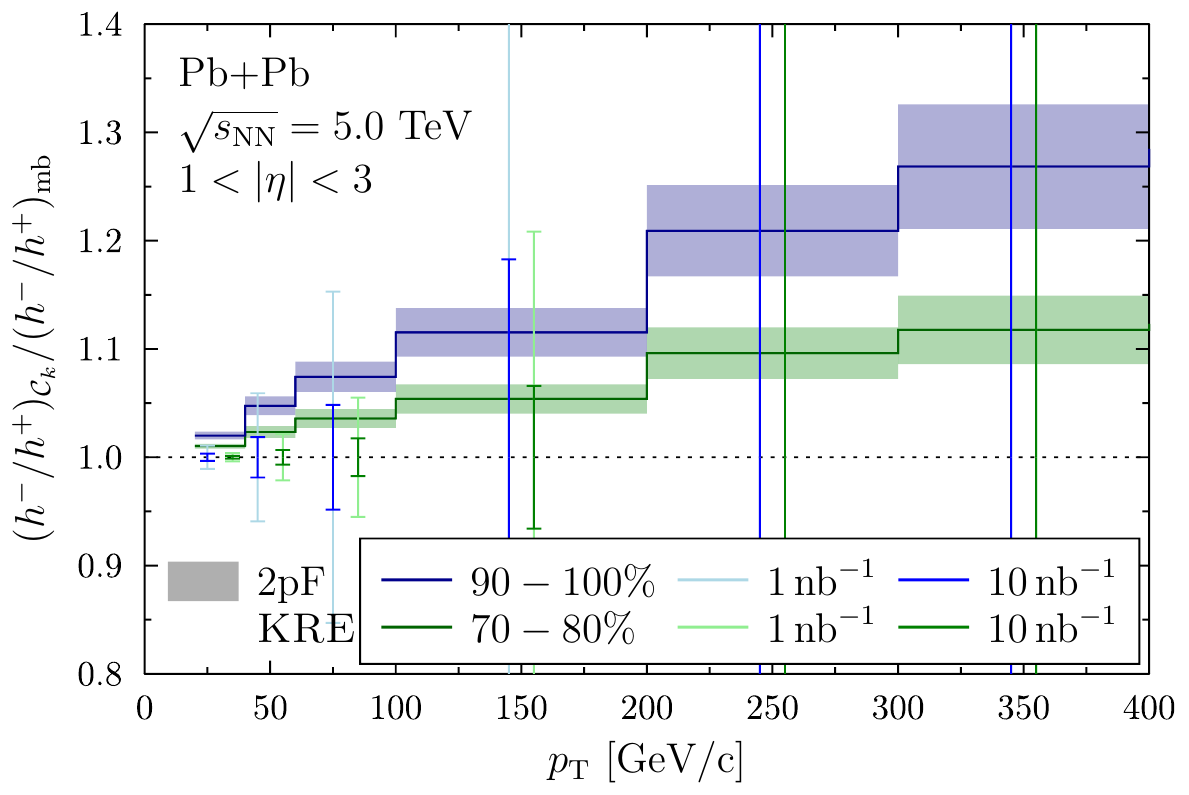}
\caption{The expected statistical precision for $1~\mathrm{nb}^{-1}$ (light vertical bars) and for $10~\mathrm{nb}^{-1}$ (darker vertical bars) nucleon--nucleon luminosity in 70--80~\% (green) and 90--100~\% (blue) centrality classes. The shaded boxes show the 2pF uncertainty as in the lower panels of Fig.~\ref{fig:RnegPosHadr}. The Kretzer FFs have been used.}
\label{fig:stats}
\end{figure}

To estimate the achievable experimental precision for the $h^-/h^+$ ratios discussed above, we multiply the cross sections from Eq.~(\ref{eq:masterformula}) by the nominal Pb-Pb nucleon--nucleon luminosity of $\mathcal{L}_{\rm nn} = 1~\mathrm{nb}^{-1}$ (also with $\mathcal{L}_{\rm nn}=10~\mathrm{nb}^{-1}$ which would correspond to the luminosity targeted after the LHC Long Shutdown 2 \cite{highlumi}) 
\footnote{In these rough estimates we do not consider in detail the uncertainties related e.g. to the use of different FFs, next-to-NLO corrections, choices for the fragmentation/factorization/renormalization scales, or suppression of the hadron yields in Pb+Pb relative to p+p baseline.
}.
From the resulting number of events $N$ we compute the relative statistical uncertainty by $1/\sqrt{N}$. To have better statistics, we consider here the rapidity bin $1 < |\eta| < 3$ (excluding the mid-rapidity to have a larger effect from the neutron skin) and suitably wide $p_{\rm T}$ intervals. The results are shown in Fig.~\ref{fig:stats} where the total statistical uncertainty follows from combining the statistical uncertainties for $h^+$ and $h^-$ quadratically. As can be seen from Fig.~\ref{fig:stats}, the realistically measurable region is $p_{\rm T} < 200~\mathrm{GeV/c}$ for the 70--80~\% bin and $p_{\rm T} < 100~\mathrm{GeV/c}$ for the 90--100~\% bin. We would expect that systematic uncertainties are small in these ratios like they are e.g. in the case of W charge asymmetry.

\section{Summary and Outlook}

We have studied the impact of the NS effect to direct-photon and charged-hadron production in Pb+Pb collisions at the LHC. In the case of photon production the NS effect has a 5--10~\% impact on $R_{\rm PbPb}^{\gamma}$ though the uncertainties in the nPDFs and their spatial dependence are of the same order or even larger than the expected effect. With $R_{\rm CP}^{\gamma}$ some of the nPDF uncertainties cancel out making the NS effect more transparent. Also, going to larger rapidities decreases the nPDF uncertainties, but still the smallness of the NS effect and the ambiguities due to the centrality dependence of the nPDFs makes the direct-photon production a challenging observable to study the NS effect.

A more promising observable is the ratio between negatively and positively charged high-$p_{\mathrm {T}}$ hadrons, for which we find up to 10~\% effects in the statistically relevant $p_{\mathrm {T}}$ region. In this case, the spatial dependence of the nPDFs cancel out very efficiently and, in general, the NS effect has a more pronounced impact than in the case of direct photons. The downsides here are the sensitivity to the applied fragmentation functions and, towards smaller $p_{\mathrm {T}}$, possible final-state modifications due to the produced strongly interacting medium. The first one can be cured by normalizing the ratio with the minimum bias result, but for a more detailed study of the latter, further modelling would be required. However, as discussed, there are indications that the final-state effects may largely disappear when considering particle ratios like the ones we have done here and, after all, the disparity between the amount of initial-state up and down quarks should strongly correlate with the balance of produced negatively and positively charged hadrons, irrespectively of the exact way the produced hard partons hadronize.

We hope that in near future the NS effect could provide an additional handle to control the centrality classification and help to bridge the theoretical and experimental centrality definitions. As a further prospect, we plan to study the NS effect in the future high-luminosity lepton-ion colliders.


\begin{acknowledgements}
I.~H. has been supported by the MCnetITN FP7 Marie Curie Initial Training Network, Contract PITN-GA-2012-315877 and has received funding from the European Research Council (ERC) under the European Union’s Horizon 2020 research and innovation programme (Grant Agreement No 668679). This research was supported by the European Research Council Grant HotLHC ERC-2011-StG-279579 and by Xunta de Galicia (Conselleria de Educacion)--H.~P. is part of the Strategic Unit AGRUP2015/11.
\end{acknowledgements}


\end{document}